# Optical readout of the chemical potential of two-dimensional electrons

Zhengchao Xia[1*], Yihang Zeng[2*], Bowen Shen[1], Roei Dery[2], Kenji Watanabe[3], Takashi Taniguchi[3], Jie Shan[1,2,4**], Kin Fai Mak[1,2,4**]

[1]School of Applied and Engineering Physics, Cornell University, Ithaca, NY, USA.
[2]Department of Physics, Cornell University, Ithaca, NY, USA.
[3]National Institute for Materials Science, Tsukuba, Japan.
[4]Kavli Institute at Cornell for Nanoscale Science, Ithaca, NY, USA.

[*]These authors contributed equally.
[**]Email: jie.shan@cornell.edu; kinfai.mak@cornell.edu

**The chemical potential ($\mu$) of an electron system is a fundamental property of a solid. A precise measurement of $\mu$ plays a crucial role in understanding the electron interaction and quantum states of matter. However, thermodynamics measurements in micro and nanoscale samples are challenging because of the small sample volume and large background signals. Here, we report an optical readout technique for $\mu$ of an arbitrary two-dimensional (2D) material. A monolayer semiconductor sensor is capacitively coupled to the sample. The sensor optical response determines a bias that fixes its chemical potential to the band edge and directly reads $\mu$ of the sample. We demonstrate the technique in AB-stacked $MoTe_2$/$WSe_2$ moiré bilayers. We obtain $\mu$ with DC sensitivity about 20 $\mu eV/\sqrt{Hz}$, and the compressibility and interlayer electric polarization using AC readout. The results reveal a correlated insulating state at the doping density of one hole per moiré unit cell, which evolves from a Mott to a charge-transfer insulator with increasing out-of-plane electric field. Furthermore, we image $\mu$ and quantify the spatial inhomogeneity of the sample. Our work opens the door for high spatial and temporal resolution measurements of the thermodynamic properties of 2D quantum materials.**

**Main**
Currently, the chemical potential or the compressibility of 2D electron systems is probed by capacitance measurements [1-6], a single-electron transistor [7-10], or a capacitive graphene sensor in double layer structures [11-13]. These measurements are all based on electrical readout and limited in speed. They cannot be spatially resolved except for the scanning single-electron transistor. In contrast, optical readout, if realized, has several potential advantages. First, it is fast and compatible with multi-channel detection that is suitable for imaging. Second, high spatial resolution (100's of nanometers) and high temporal resolution (10's-100's of femtoseconds when combined with ultrafast optical pulses) are possible. Third, in contrast to the scanning single-electron transistor, it can access samples embedded in dual-gated device structures. Finally, optical readout is also compatible with simultaneous spectroscopy measurements on 2D electron systems. In this study, we demonstrate optical readout of the chemical potential in 2D quantum materials. We report a DC sensitivity of about 20 $\mu eV/\sqrt{Hz}$, on par with that of the established techniques. We also show detection of the compressibility and the out-of-plane electric polarization by AC readout and imaging of the chemical potential.

**Working principle**
The technique employs a monolayer semiconductor as a sensor whose optical response sharply depends on the chemical potential [14, 15]. The sensor is capacitively coupled to the sample, similar to the case of the graphene sensor for electrical readout of $\mu$ (Ref. [11-13]). A bias is applied to lock the sensor chemical potential to a reference point according to the optical response through a feedback circuit. The bias directly reads $\mu$ of the sample.

Figure 1a illustrates an implementation of the idea based on a dual-gated device. The sensor is inserted between the sample and the bottom gate. The gates, sensor and sample form parallel-plate capacitors in series. We electrically ground the sample and control the top gate bias ($V_{tg}$), the sensor bias ($V_s$) and the bottom gate bias ($V_{bg}$). The top gate and the sensor serve as two independent gates that tune the doping density ($n$) in and the electric field ($E$) perpendicular to the sample. The bottom gate voltage fixes the sensor chemical potential at a reference point. We can relate the sample chemical potential to the gate voltages, $\mu/e = \left(1 + \frac{C_{bg}}{C_s}\right) V_s - \frac{C_{bg}}{C_s} V_{bg}$, up to a constant energy shift (Methods). Here $e$ is the elementary charge, $C_{bg}$ and $C_s$ are the bottom-gate-to-sensor and sensor-to-sample geometrical capacitances, respectively, and their ratio can be independently calibrated in experiment (Extended Data Fig. 2). A natural reference point for the sensor chemical potential is its conduction or valence band edge. This can be accurately probed optically because monolayer semiconductors interact strongly with light and the fundamental exciton resonance is extremely sensitive to doping [14, 15]. High detection sensitivity for the chemical potential is therefore possible.

In addition to the DC readout modality described above that measures $\mu$, the technique also allows differential AC measurements for other thermodynamic properties. For instance, $\left(\frac{\partial \mu}{\partial n}\right)_E$ is the inverse compressibility up to a factor of $n^2$; $\left(\frac{\partial \mu}{\partial E}\right)_n$ is related to the electric polarization density $P$ through the Maxwell's relation, $\left(\frac{\partial \mu}{\partial E}\right)_n = -\left(\frac{\partial P}{\partial n}\right)_E$. We superpose small AC biases of distinct frequencies on the three DC biases, $V_{tg}$, $V_{bg}$ and $V_s$, and detect the corresponding optical modulations using the lock-in technique. At the same time, the DC optical output is used to fix the sensor chemical potential at its reference point. The inverse compressibility ($\left(\frac{\partial \mu}{\partial n}\right)_E$) and the differential electric polarization density ($\left(\frac{\partial P}{\partial n}\right)_E$) can be obtained as linear combinations of the three AC optical outputs (Methods). Other properties, such as the magnetization [16] and entropy [17, 18], can be similarly measured by modulating the magnetic field and temperature, respectively.

**The chemical potential and compressibility**
We demonstrate the technique in AB-stacked MoTe$_2$/WSe$_2$ moiré bilayers, where rich correlated electronic phases have been observed in recent transport and optical studies [19-22]. The two transition metal dichalcogenide (TMD) monolayers form a triangular moiré lattice of period around 5 nm [19-22], corresponding to a moiré density of $n_M \approx 5 \times 10^{12}$ cm$^{-2}$. *Ab initio* calculations have shown that the first and second moiré valence bands are made of localized Wannier orbitals in the MoTe$_2$ and WSe$_2$ layers, respectively [23]. They each form a triangular lattice, and together, a honeycomb lattice [23-26]. The sublattice potential difference can be controlled by an out-of-plane electric field, and electric-field-induced band inversion and topological phase transitions have been reported [19, 21]. Of particular interest are the chemical potential changes around hole doping density $n = n_M$ (or moiré unit cell filling factor $\nu = 1$), where the electron interaction can drive the system at half-band filling into a correlated insulator.

We fabricate the device using the layer-by-layer transfer technique [27] (Methods). We use a WSe$_2$ monolayer as the sensor and hexagonal boron nitride (hBN) as the dielectric in the capacitors. The hBN thickness is 5, 2.7 and 5.5 nm, respectively, in the top gate, between the sample and sensor, and in the bottom gate (Fig. 1a). Figure 1b is the reflectance spectrum of the sensor as a function of $V_{bg}$, which controls the sensor chemical potential. The other two biases are fixed. A prominent neutral exciton resonance is observed near 1.73 eV. It rapidly loses the spectral weight upon electron doping and evolves into a much weaker (blue-shifted) repulsive polaron [28]. The reflection of the neutral exciton (integrated from 1.727 - 1.731 eV) sharply depends on $V_{bg}$ (Fig. 1c). We choose the steepest point as the reference for maximal sensitivity. For each measurement, $V_{bg}$ is tuned through a feedback loop to lock reflection at the reference point. In this arrangement, the sensor chemical potential is fixed at the conduction band edge, and the sample chemical potential is at $eV_s$ below it (Fig. 1d). We define $\mu = 0$ at the moiré band edge. Figure 1e,f characterize the detection sensitivity of $\mu$. Four hundred measurements are performed consecutively with fixed top gate and sensor biases. The integration time for each is 0.13 s. The histogram of the measured $\mu$ after numerically removing the long-term drift shows a standard deviation of about 65 μeV. It corresponds to a DC sensitivity of about 20 μeV/√Hz; it is about 3 times above the shot-noise limit because the feedback loop stops when the reflectance reaches below 0.2 % difference from the reference point (Methods).

Figure 2a-c show the filling dependence of the chemical potential of the moiré bilayer at 3 K for electric field $E \approx 0.56$, 0.65 and 0.70 V/nm, respectively. Figure 2d-f are the corresponding inverse compressibility from the AC measurement. In all cases, $\mu$ increases rapidly upon doping and a peak is observed around $\nu = 0$. The filling factor can be determined from the gate voltages or from the known filling factors for the correlated insulating states [19] (Methods). It is not expected to be accurate near $\nu = 0$ because of the poor electrical contacts (when the sample doping density is very low) and the associated nonlinear gating effect. With further increase of $\nu$, we observe a generally decreasing $\mu$ or negative compressibility. Similar negative compressibility has been reported in MoSe$_2$/WS$_2$ moiré bilayers [29]. It is a manifestation of the strong electron interaction in flat band systems [3, 6, 29]. The interaction effect and negative compressibility are significantly reduced at higher doping densities ($\nu > 1$). Furthermore, across $\nu = 1$, we observe an abrupt increase of $\mu$, or equivalently, a peak in $\left(\frac{\partial \mu}{\partial \nu}\right)_E$ for the two low electric fields (Fig. 2a, b). This reveals an incompressible state at $\nu = 1$. The charge gap is determined to be $\Delta \approx 25$ and 8 meV for $E \approx 0.56$ and 0.65 V/nm, respectively, from the jump amplitude of $\mu$. We also observe a weaker incompressible state at $\nu = \frac{1}{3}$ under all three electric fields and at $\nu = \frac{2}{3}$ under lower electric fields (not shown).

**The interlayer electric polarization**
The incompressible state observed at $\nu = 1$ is a correlated insulator [19, 20]. To better understand its nature, we probe $\left(\frac{\partial \mu}{\partial E}\right)_n$ through the AC measurement in Fig. 3a. The quantity reflects the change in the interlayer electric polarization per injected particle. Its sign indicates in which TMD layer the injected hole resides, i.e. the interlayer charge distribution. We can also obtain $P(v)$ in Fig. 3b via integration, $P/n_M = -\int_0^v dv \left(\frac{\partial \mu}{\partial E}\right)_n$, by requiring $P = 0$ at $v = 0$. This quantity is proportional to the density imbalance between the two TMD layers and the renormalized dielectric function. The behavior of the insulating state under $E \approx 0.56$ V/nm and 0.65 V/nm is very different. Under the small field, $\left(\frac{\partial P}{\partial E}\right)_n$ evolves smoothly across $v = 1$, but under the large field it changes abruptly from a positive to a negative value. Consistently, under the small field $P$ shows a smooth doping dependence, but under the large field it shows a pronounced negative peak at $v = 1$ followed by a rapid increase in the opposite direction.

Our result indicates that under the small field the doped holes reside only in one TMD layer ($MoTe_2$) as the chemical potential sweeps through the incompressible state at $v = 1$, but under the large field the additional holes are doped into the second TMD layer ($WSe_2$). This picture is further verified by the optical spectroscopy measurement on the moiré bilayer. Figure 3c is the doping dependence of the reflectance contrast spectrum of the sample near the $WSe_2$ fundamental exciton resonance (which occurs at a lower energy than in the $WSe_2$ sensor layer due to the stronger dielectric screening effect in the moiré bilayer [30]). Under the small field, the neutral exciton resonance survives for $v > 1$, but under the large field it abruptly turns into the polaron feature beyond $v = 1$, signifying hole doping in the $WSe_2$ layer.

The incompressible state at $v = 1$ is therefore consistent with a Mott insulator under $E \approx 0.56$ V/nm and a charge-transfer insulator under $E \approx 0.65$ V/nm. The corresponding band diagrams are illustrated in the inset of Fig. 2a, b. The purple and red curves denote the $MoTe_2$ Hubbard bands and the $WSe_2$ dispersive moiré band, respectively [22, 25]. Under $E \approx 0.56$ V/nm, the $WSe_2$ moiré band is below the upper Hubbard band. The chemical potential jump at $v = 1$ is equal to the gap between the $MoTe_2$ Hubbard bands, and the gap is stabilized by the strong on-site Coulomb repulsion [31]. On the other hand, under $E \approx 0.65$ V/nm the $WSe_2$ moiré band is between the $MoTe_2$ Hubbard bands. The chemical potential jump at $v = 1$ is the charge-transfer gap between the $MoTe_2$ lower Hubbard band and the $WSe_2$ moiré band [22]. It is smaller than the Mott gap and continuously tunable by the electric field (Fig. 4). Under even higher fields (e.g. $E \approx 0.70$ V/nm), the $WSe_2$ moiré band can overlap with the $MoTe_2$ lower Hubbard band, the $v = 1$ charge gap closes, and the system becomes a semimetal (Fig. 2c).

**Imaging the thermodynamic equation of state**
Finally, we demonstrate imaging $\mu$ of the sample as a function of $v$ and $E$, a thermodynamic equation of state. Figure 4a shows an optical image of the device; the black line encircles the region of interest, where the sensor is relatively uniform. We apply a wide-field-of-view illumination using a narrow-band light source that covers the fundamental exciton resonance of the sensor. The reflected light from the device is sent to a charge-coupled device (CCD) for imaging. We characterize the sensor response and measure $\mu$ for each pixel using the DC readout method described above for a single point (see Methods for details). Since scanning the beam or the sample is not required, fast chemical potential imaging can be achieved. The spatial resolution is diffraction-limited and is about 1 µm (Extended Data Fig. 3).

Figure 4b illustrates the measured $\mu(\nu, E \approx 0.56 \text{ V/nm})$ at representative locations denoted by dots of the same color in Fig. 4a. Clear spatial inhomogeneity is observed. The charge gap determined from the chemical potential jump around $\nu = 1$ is mapped in Fig. 4e. It varies by a few times throughout the sample. In addition, the filling factor at which the chemical potential jump occurs, also deviates from '$\nu = 1$' at different locations. This reflects the spatial variations of the moiré density $n_M$ (Fig. 4d), which are about ± 10% throughout the sample. We further probe the local band alignment by measuring the electric-field dependence of the charge gap at $\nu = 1$ under large electric fields (Fig. 4c). The decreasing gap size with the increasing field is consistent with the behavior of a charge-transfer insulator. All curves (locations) show a slope of about $0.3\ e \cdot nm$, which agrees well with the reported interlayer dipole moment in MoTe$_2$/WSe$_2$ moiré bilayer [32]. The spatial variations of the critical electric field $E_c$ at which the charge gap vanishes are shown in Fig. 4f. We observe weak correlation between Δ and $n_M$, but a negative correlation between Δ and $E_c$ (Extended Data Fig. 4).

Our result suggests that the twist angle disorder in the device, which causes variations in $n_M$, is not significant. This is typical for moiré heterobilayers, in which the moiré density is largely determined by the lattice mismatch of the two monolayers [31]. The observed strong variations in Δ are likely dominated by the random variations in the built-in electric field perpendicular to the sample plane; this changes $E_c$ randomly across the sample and the measured Δ at a fixed applied electric field. The origin for such built-in electric field disorder is currently unclear. It could come from the Coulomb disorder in the device, trapped air bubbles from the device fabrication process and/or unintentional strain disorder in the sample [33]. Further investigations are required to better understand the built-in electric field disorder to achieve more homogeneous devices.

**Conclusion**
We have demonstrated optical readout of the chemical potential of 2D materials using a semiconductor TMD sensor. The strong light-matter interaction and the doping-sensitive exciton resonance in the sensor enable chemical potential readout with high sensitivity. In addition to the chemical potential and compressibility, the technique allows the measurement of the electric polarization and other thermodynamic properties by applying the Maxwell's relations. The optical readout method further allows fast imaging of the thermodynamic equation of state and analysis of the spatial inhomogeneity of the correlated electron state. The technique paves the path for discovery of new quantum states of matter in 2D materials.

**Methods**
**Device fabrication**
We fabricated the dual-gated MoTe$_2$/WSe$_2$ moiré devices (Fig. 1a) using the reported layer-by-layer dry transfer method [27]. In short, we exfoliated monolayer WSe$_2$ and MoTe$_2$, multi-layer hBN and graphite onto silicon substrates with a 300 nm silicon dioxide layer; we identified the flakes under an optical microscope. To angle-align the two TMD layers, we performed angle-resolved optical second-harmonic generation (SHG) to identify their relative crystal axis orientation before stacking into the desired sequence [34, 35]. We also inserted a monolayer WSe$_2$ sensor in between the MoTe$_2$/WSe$_2$ moiré and the graphite bottom gate electrode; the sensor is not angle-aligned with the moiré layer and is separated from the moiré layer by a thin hBN spacer (with thickness $d_s \approx 2.7$ nm). The sample and sensor are separately contacted by few-layer graphite electrodes. In order to apply a large electric field (up to about 1 V nm$^{-1}$) perpendicular to the sample plane, we used hBN flakes with thicknesses $d_{tg} = 5$ nm and $d_{bg} = 5.5$ nm for the top and bottom gate dielectrics, respectively. In this device geometry, the top gate bias $V_{tg}$ and the sensor bias $V_s$ can independently tune the doping density and electric field in the moiré bilayer. The doping density can be calibrated based on the known filling factors of the correlated insulating states and the moiré density. We define the electric field as $E = [(V_s + \phi)/d_s - V_{tg}/d_{tg}]/2$, where $\phi = 1.37$ V is the work function difference between the monolayer WSe$_2$ sensor and the graphite gate electrode.

**Optical measurements**
We performed chemical potential sensing measurements on devices mounted in a closed-cycle optical cryostat (Attocube, attoDRY2100). For DC measurements, we illuminated our devices by a LED light source with spectral range 710-750 nm and total power less than 0.8 nW; we collected the reflected light from the device using a large numerical aperture objective (N.A. = 0.8) and sent the light to a spectrometer equipped with a charge-coupled device (CCD), which recorded the reflected photon counts from the sensor as a function of the wavelength. We then integrated the photon counts over a spectral window (716.3-718.0 nm) that covers the neutral exciton resonance of the sensor, and obtained a response curve such as the one shown in Fig. 1c. For chemical potential readout, we first picked a reference point in the response curve with the maximum slope (for optimal sensitivity). We then measured the reflected photon count from the sensor at each filling factor and electric field and compared the count to the reference point. To complete the feedback loop, we adjusted the bottom gate voltage $V_{bg}$ to reduce the count difference from the reference point to below 0.2 %. We readout the value of $V_{bg}$ to obtain the chemical potential of the sample (see Electrostatics below).

For AC measurements, we modulated the top gate, bottom gate and sensor bias voltages at 57.777, 17.777 and 37.777 Hz, respectively. We probed the neutral exciton response of the sensor using a 717.7 nm laser (M Squared SolsTiS) with incident power less than 200 nW, and measured the reflected light intensity $I$ from the sensor using a silicon avalanche photodiode (APD). We used lock-in amplifiers to measure the demodulated APD output signals $\left(\frac{\partial I}{\partial V_{tg}}\right)$, $\left(\frac{\partial I}{\partial V_{bg}}\right)$ and $\left(\frac{\partial I}{\partial V_s}\right)$ at the respective frequencies. We also measured the DC output from the APD using a multi-meter and low-pass filter (Fig. 1a), and used the same feedback loop as in DC measurements to keep the DC output near the fixed reference point of the sensor response curve. As we will show in the section Electrostatics below, this will allow us to obtain the quantities $\left(\frac{\partial \mu}{\partial n}\right)_E$ and $\left(\frac{\partial \mu}{\partial E}\right)_n$.

**Imaging the chemical potential**
To image the chemical potential of the sample, we illuminated the region of interest using a 730-nm light-emitting diode (LED) under wide-field-of-view illumination. The light spectrum is filtered to include 1.722 - 1.741 eV, which covers the spatially inhomogeneous neutral exciton resonance of the sensor throughout the region of interest. The reflected light from the sensor was sent to a charge-coupled-detector (CCD) for imaging. We first obtained the local sensor optical response, similar to the one in Fig. 1c, for each pixel of the image. The response deviates slightly from pixel to pixel due to the spatial inhomogeneity of the sensor (variations in the fundamental exciton energy and unintentional doping density). We chose a representative pixel 'R', and for a given $(\nu, E)$ we applied bottom-gate bias, $V_{bg}^R$, to fix the photon count at the reference point via a feedback loop. To determine $\mu$ in other pixels, we measured the optical reflection at four other bottom-gate biases around $V_{bg}^R$ (including $V_{bg}^R$) and shifted the local sensor response in $V_{bg}$ to fit the five data points. Examples are shown in Extended Data Fig. 5.

**Electrostatics**

We model the dual-gated device as three parallel-plate capacitors in series (Extended Data Fig. 1) with geometrical capacitance $C_{tg} = \frac{\varepsilon}{d_{tg}}$, $C_{bg} = \frac{\varepsilon}{d_{bg}}$ and $C_s = \frac{\varepsilon}{d_s}$ for the top gate, bottom gate and between the sample and sensor, respectively. Here $\varepsilon$ is the dielectric constant of hBN and $d_i$ is the hBN thickness in the top gate ($i = tg$), the bottom gate ($i = bg$) and between the sample and sensor ($i = s$). The sample is grounded. Three biases, $V_{tg}$, $V_s$ and $V_{bg}$, are applied to the top gate, sensor and the bottom gate, respectively. The resultant out-of-plane electric fields are $E_{tg}$, $E_s$ and $E_{bg}$, respectively, between the top gate and sample, between the sample and sensor, and between the sample and bottom gate. The chemical potential and doping density are $(\mu, n)$ in the sample and $(\mu_s, n_s)$ in the sensor.

We relate the charge density in the sample and sensor to the electric fields in the parallel-plate capacitors using the appropriate boundary conditions [36]

$$\begin{cases} ne = \varepsilon(E_s - E_{tg}) \\ n_s e = \varepsilon(E_{bg} - E_s) \end{cases}. \tag{1}$$

The electrochemical potentials ($V_{tg}$, $V_s$ and $V_{bg}$) are related to the chemical and electrostatic potentials as

$$\begin{cases} eV_{tg} = \mu - eE_{tg}d_{tg} \\ eV_s = \mu - \mu_s + eE_s d_s \\ eV_{bg} = eV_s + \mu_s + eE_{bg}d_{bg} \end{cases}. \tag{2}$$

Solving equation (1) and (2) gives

$$\frac{\mu}{e} = \left(1 + \frac{C_{bg}}{C_s}\right)V_s - \frac{C_{bg}}{C_s}V_{bg} + \left(1 + \frac{C_{bg}}{C_s}\right)\frac{\mu_s}{e} + \frac{n_s e}{C_s}. \tag{3}$$

In our experiment, $\mu_s$ and $n_s$ are kept as constants by the feedback loop. After taking into account a negative sign for holes, the hole chemical potential is given up to a constant energy shift as

$$\frac{\mu}{e} = \frac{C_{bg}}{C_s}V_{bg} - \left(1 + \frac{C_{bg}}{C_s}\right)V_s. \tag{4}$$

To relate the quantities $\left(\frac{\partial I}{\partial V_{tg}}\right)$, $\left(\frac{\partial I}{\partial V_{bg}}\right)$ and $\left(\frac{\partial I}{\partial V_s}\right)$ from AC measurements to $\left(\frac{\partial \mu}{\partial n}\right)_E$ and $\left(\frac{\partial \mu}{\partial E}\right)_n$, we note that only $V_{tg}$ and $V_s$ are independent variables that tune the sample chemical potential $\mu(V_{tg}, V_s) = \mu(\nu, E)$; $V_{bg}$ is determined from the feedback loop and is not an independent variable. We have

$$\begin{cases} \left(\frac{\partial \mu}{\partial V_{tg}}\right) = -\frac{C_{bg}}{C_s}\left(\frac{\partial I}{\partial V_{tg}}\right)\left(\frac{\partial I}{\partial V_{bg}}\right)^{-1} \\ \left(\frac{\partial \mu}{\partial V_s}\right) = -\frac{C_{bg}}{C_s}\left(\frac{\partial I}{\partial V_s}\right)\left(\frac{\partial I}{\partial V_{bg}}\right)^{-1} - \left(1 + \frac{C_{bg}}{C_s}\right) \end{cases}. \quad (5)$$

Using the linear expressions that connect $\nu$ and $E$ to $V_{tg}$ and $V_s$, we can obtain $\left(\frac{\partial \mu}{\partial n}\right)_E$ and $\left(\frac{\partial \mu}{\partial E}\right)_n$ using the above equations and the following expressions

$$\begin{cases} \left(\frac{\partial \mu}{\partial n}\right)_E = \frac{e}{2C_{tg}}\left(\frac{\partial \mu}{\partial V_{tg}}\right) + \frac{e}{2C_s}\left(\frac{\partial \mu}{\partial V_s}\right) \\ \left(\frac{\partial \mu}{\partial E}\right)_n = \left(\frac{\partial \mu}{\partial V_s}\right) d_s - \left(\frac{\partial \mu}{\partial V_{tg}}\right) d_{tg} \end{cases}. \quad (6)$$

**Shot-noise-limited sensitivity**

The photon shot-noise is the square root of the photon count at the reference point $\sqrt{N}$. This contributes to a chemical potential noise $\delta\mu \approx \frac{C_{bg}}{C_s}\frac{\Delta V_{bg}}{SNR} \sim 6$ μV/√Hz, in which $SNR = \sqrt{N} \sim 2500$ √Hz is the signal-to-noise ratio of the optical measurement and $\Delta V_{bg} \sim 30$ mV is the width of the response curve for the sensor band edge (Fig. 1c). The value is about three times lower than the measured noise because the feedback loop stops when the reflectance reaches below 0.2 % difference from the reference point. We can reduce the noise level by decreasing the ratio $\frac{C_{bg}}{C_s}$ in our device design, increasing $SNR$ in the optical measurement, and reducing the sensor edge width $\Delta V_{bg}$ by minimizing the disorder level in the sensor.

**Acknowledgement**
We thank Andrew Pierce for the helpful suggestions on AC measurements.

# Figures

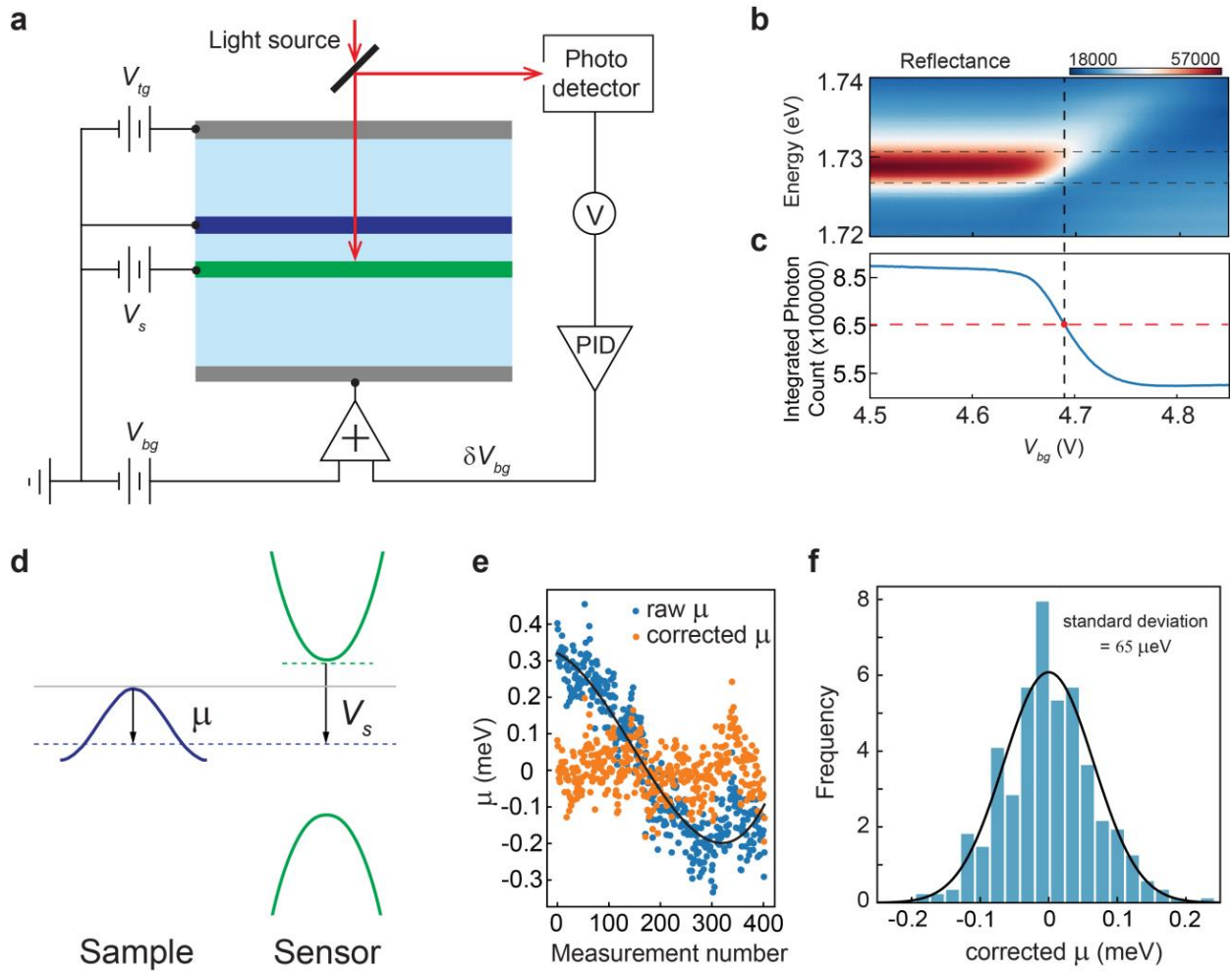

**Figure 1 | Optical readout of the chemical potential. a,** Schematic setup for chemical potential sensing in a dual-gated device of AB-stacked MoTe$_2$/WSe$_2$ moiré bilayer. The top gate and sensor serve as two independents gates to the sample; the bottom gate sets the sensor at its reference point through a feedback loop and optical readout. See Methods for details. **b,** Bottom gate voltage dependence of the reflectance spectrum near the neutral exciton resonance of the WSe$_2$ sensor. The exciton resonance is rapidly quenched upon electron doping near 4.69 V. **c,** Integrated photon counts over the spectral window in **b** (horizontal dashed lines) as a function of the bottom gate voltage. The vertical dashed line marks the reference point of the sensor. **d,** Schematic band alignment of the device. The purple and green curves label the electronic bands of the sample and sensor, respectively. The dashed lines denote the Fermi level of each layer. The arrows label $V_s$ and $\mu$. **e,** 400 measurements of the chemical potential taken continuously in time with fixed $V_{tg}$ and $V_s$. The blue dots are the raw data; the black curve is a third-order polynomial fit to the raw data; and the orange dots are the difference between the two. **f,** Histogram of the corrected $\mu$. The black curve is a Gaussian fit to the data. The standard deviation is 65 $\mu$eV, which corresponds to a DC sensitivity about 20 $\mu$V/√Hz for an exposure time of 0.13 s.

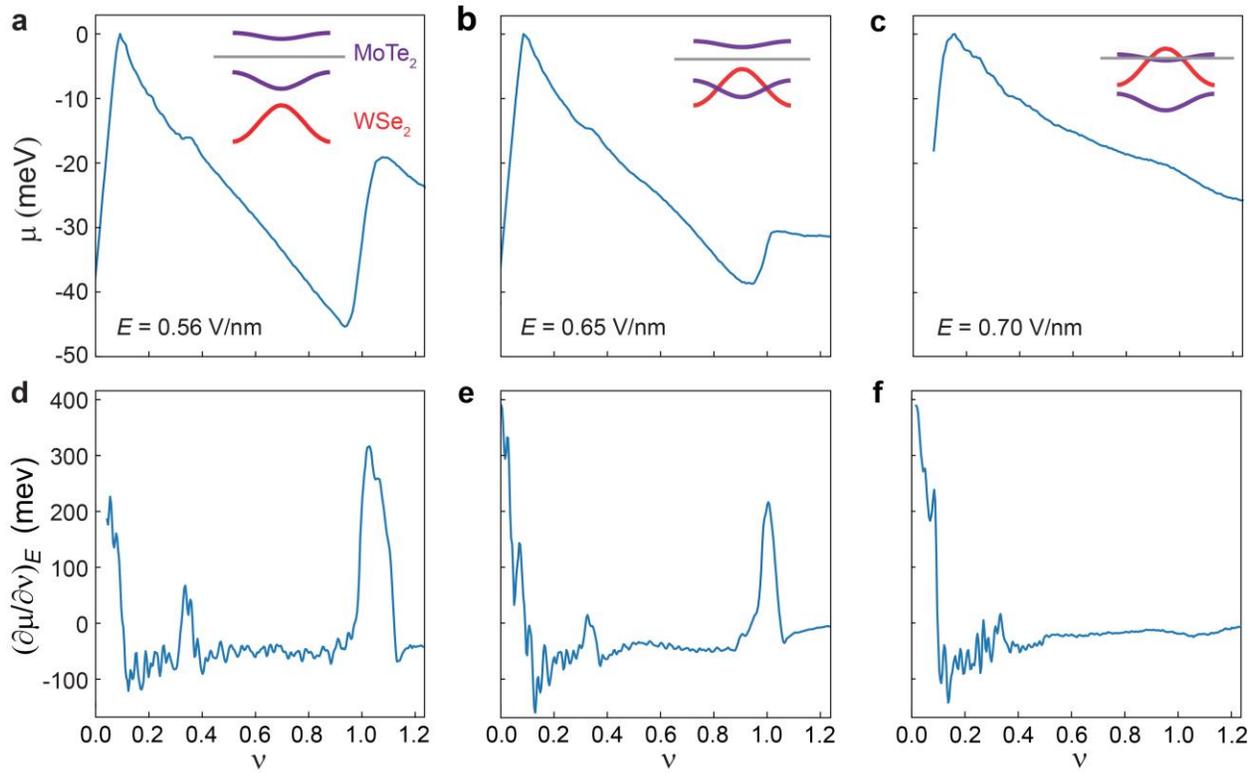

**Figure 2 | Chemical potential and compressibility. a-c,** Filling factor dependence of the hole chemical potential in AB-stacked $MoTe_2/WSe_2$ moiré bilayer at $E = 0.56$ V/nm **(a)**, 0.65 V/nm **(b)** and 0.70 V/nm **(c)**. The chemical potential peak near $\nu = 0$ marks the onset of hole doping. The chemical potential jumps at $\nu = 1/3$ and 1 correspond to incompressible states. A large negative compressibility is observed. The insets show the alignment for the lower and upper Hubbard bands of the $MoTe_2$ layer (purple) and the $WSe_2$ moiré band (orange). The system is a Mott insulator **(a)**, charge-transfer insulator **(b)** and a semimetal **(c)** at $\nu = 1$. The grey line denotes the Fermi level. **d-f,** Filling factor dependence of the inverse compressibility (from AC measurements) at the same electric fields as in **a-c**. The incompressible states are revealed as peaks at $\nu = 1/3$ and 1. The high contact resistance causes the increased noise at low doping densities.

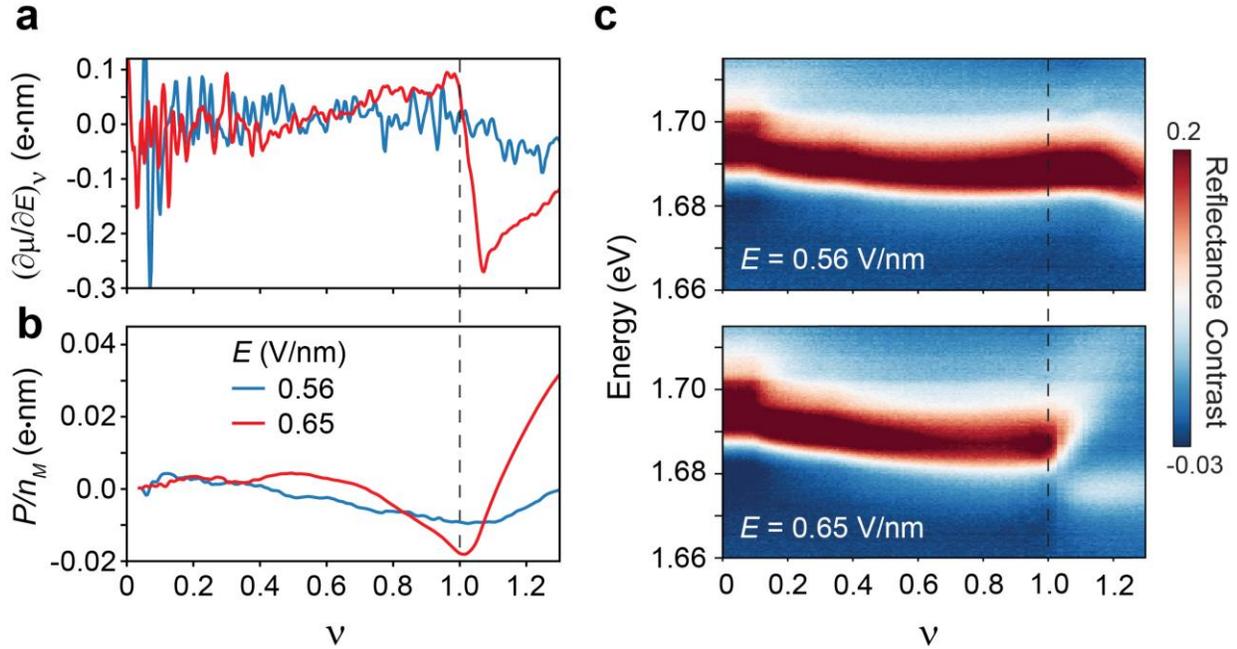

**Figure 3 | Interlayer electric polarization. a,** Filling factor dependence of $\left(\frac{\partial \mu}{\partial E}\right)_\nu$ at $E = 0.56$ V/nm (blue) and 0.65 V/nm (red). A sharp sign change at $\nu = 1$ (dashed line) is observed only for the charge-transfer insulator at $E = 0.65$ V/nm. **b,** Filling factor dependence of the interlayer electric polarization per moiré unit cell (obtained from integrating the data in **a**) at the same electric fields as in **a**. **c,** Filling factor dependence of the reflection contrast spectrum at $E = 0.56$ V/nm (upper) and 0.65 V/nm (lower). The quenched neutral exciton resonance in the moiré WSe$_2$ layer in lower panel shows the onset of hole doping into the WSe$_2$ layer beyond $\nu = 1$, consistent with the charge-transfer insulator picture. The neutral exciton resonance remains robust through $\nu = 1$ for the Mott insulator in upper panel.

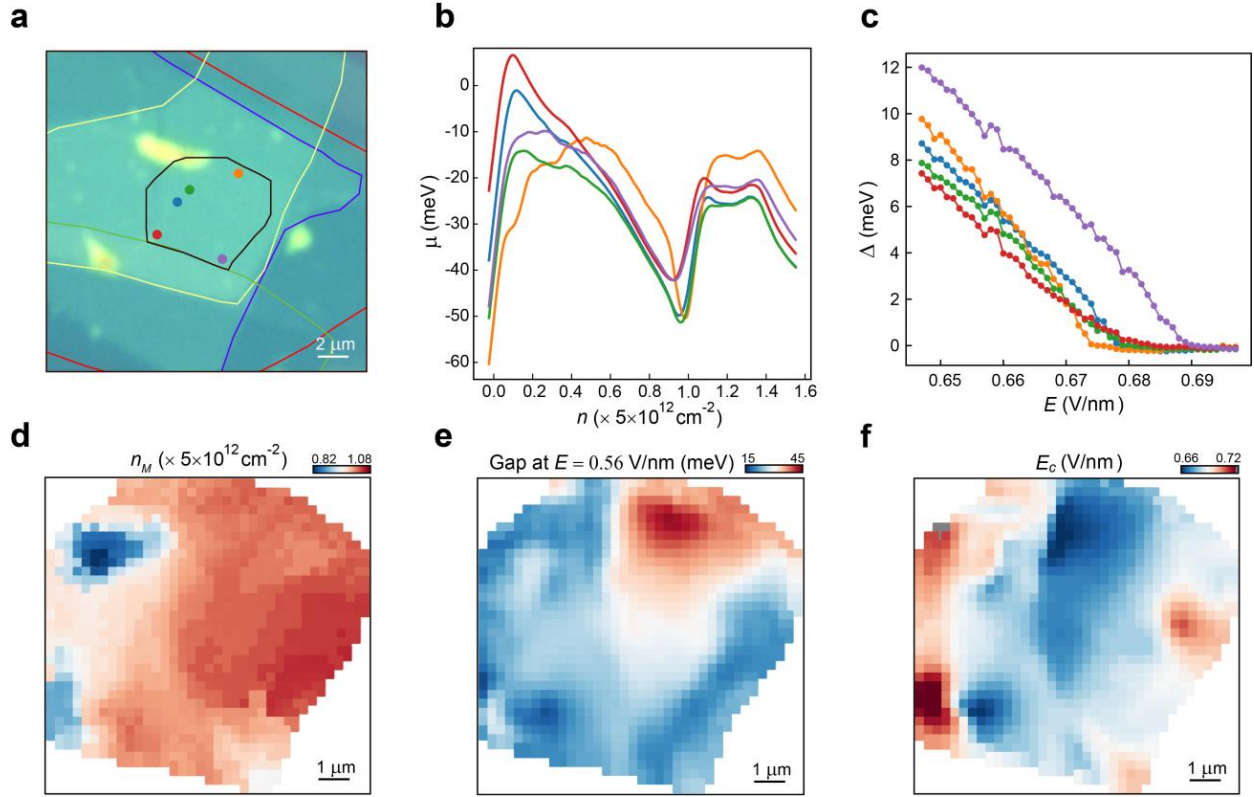

**Figure 4 | Chemical potential imaging. a,** Optical image of the dual-gated device. The blue, red and yellow lines mark the boundary of the moiré MoTe$_2$, moiré WSe$_2$ and sensor WSe$_2$, respectively; the black line encloses the region of interest for our imaging study. The colored dots mark the locations where data are shown in **b** (with the same colors). **b,** Charge density dependence of the chemical potential at a constant electric field $E = 0.56$ V/nm for the locations marked in **a**. **c.** Electric field dependence of the $\nu = 1$ charge gap at the same locations. The charge gap vanishes at the critical electric field $E_c$. Spatial inhomogeneities in both the moiré density, gap size and critical electric field are observed in **b** and **c**. **d-f,** Spatial map of the moiré density (**d**), charge gap at $E = 0.56$ V/nm (**e**) and $E_c$ (**f**). Whereas no obvious correlation is seen between **d** and **e**, the charge gap and $E_c$ show a negative correlation in **e** and **f** (see Extended Data Fig. 4).

**Extended data figures**

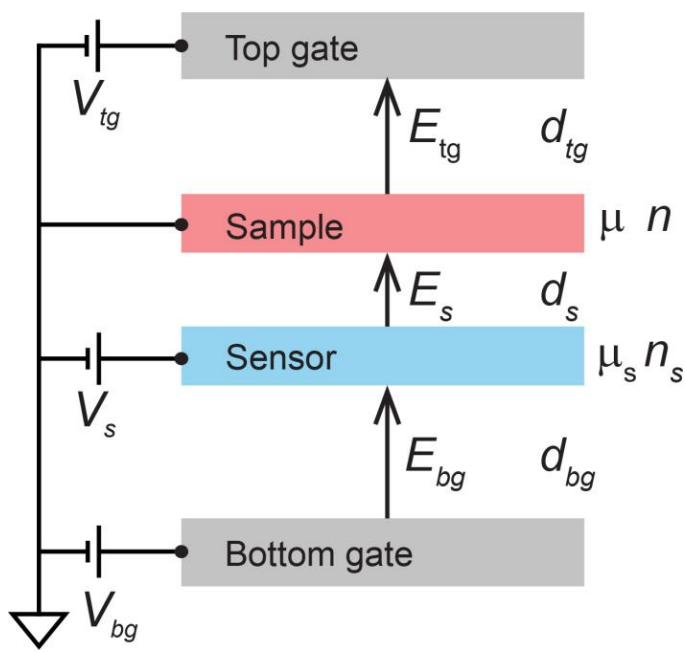

**Extended Data Figure 1 | Device schematics.** Schematic dual-gated device structure with a sensor in between the sample and bottom gate. The active layers are separated by hBN dielectrics (not shown). The electrochemical potentials, chemical potentials, electric fields and doping densities are labeled. See Methods for details.

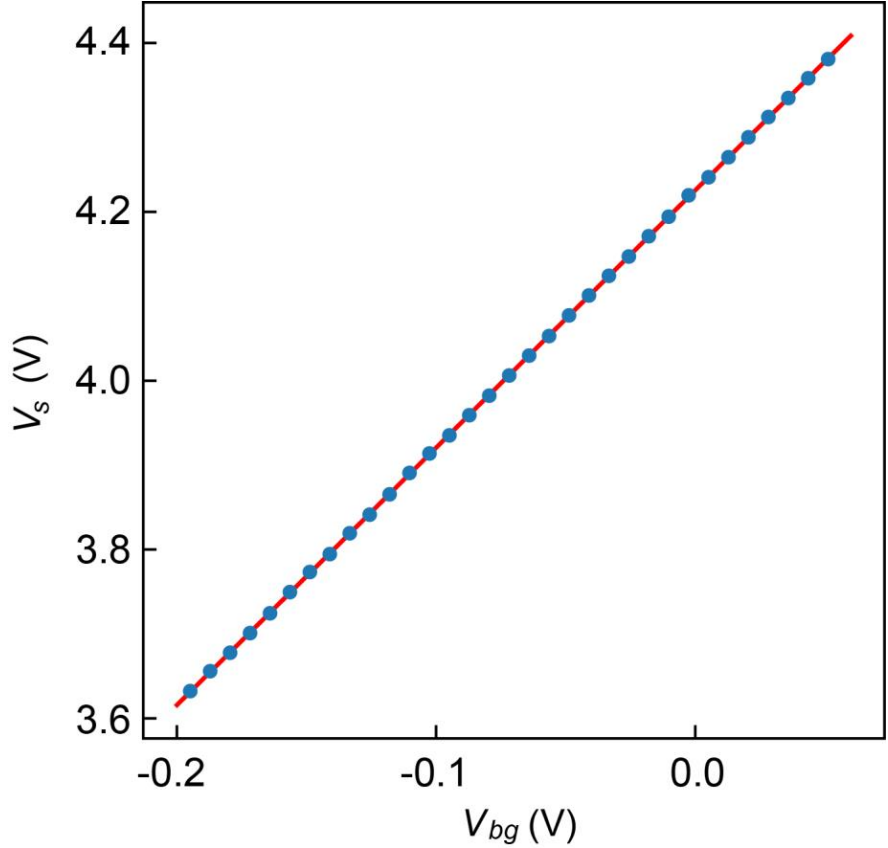

**Extended Data Figure 2 | Extracting capacitance lever arm.** To extract the capacitance lever arm, we kept the sample at charge neutrality and the sensor at the reference point (by tuning $V_{bg}$). The blue dots show the relation between $V_s$ and $V_{bg}$ when sample is tuned (by $V_{tg}$ and $V_s$) such that $(C_{tg}V_{tg} + C_s V_s)$ is kept constant; this gives $V_{bg} = \left(\frac{C_s}{C_{bg}} + 1\right)V_s + \text{constant}$. We determine the ratio $\frac{d_{bg}}{d_s} = \frac{C_s}{C_{bg}} = 2.05$ by fitting the results to a linear dependence (red line).

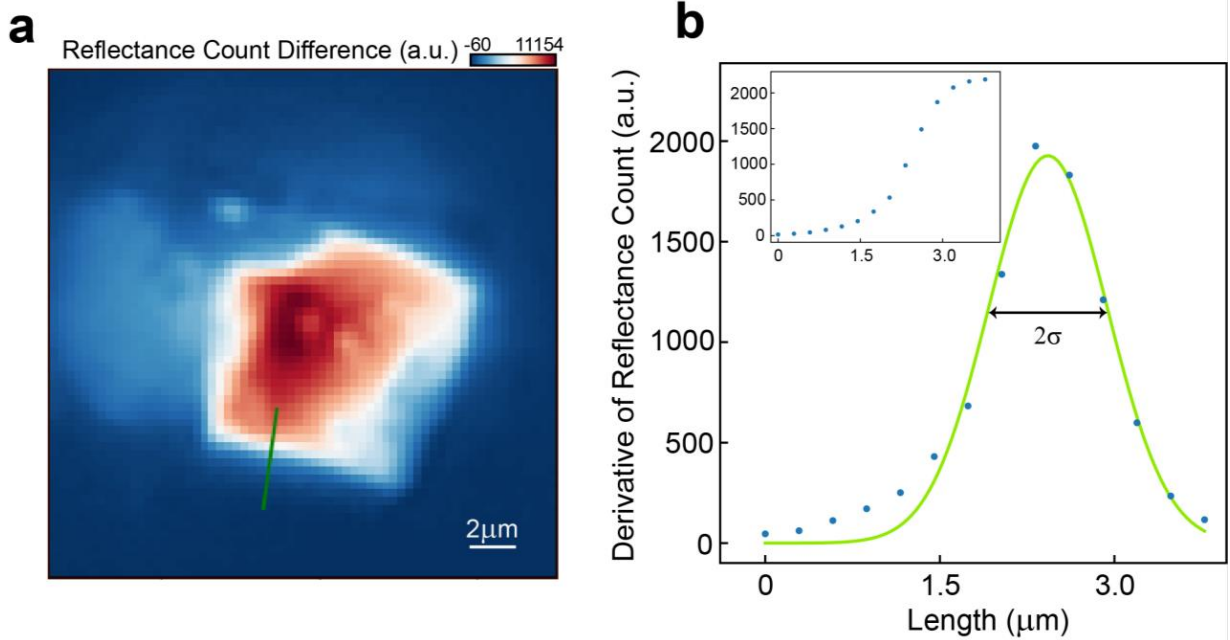

**Extended Data Figure 3 | Spatial resolution. a,** Reflection contrast image at the neutral exciton resonance of the WSe$_2$ sensor. The bright region is the region of interest where the sample covered by the sensor. **b,** Inset: Linecut of reflection count difference across a sharp sensor edge (taken at the green line in **a**). Main panel: Spatial derivative of the data in the inset (data points) and the corresponding Gaussian fit (solid line). The spatial resolution, as determined by twice of the standard deviation of the Gaussian fit, is about 1 um.

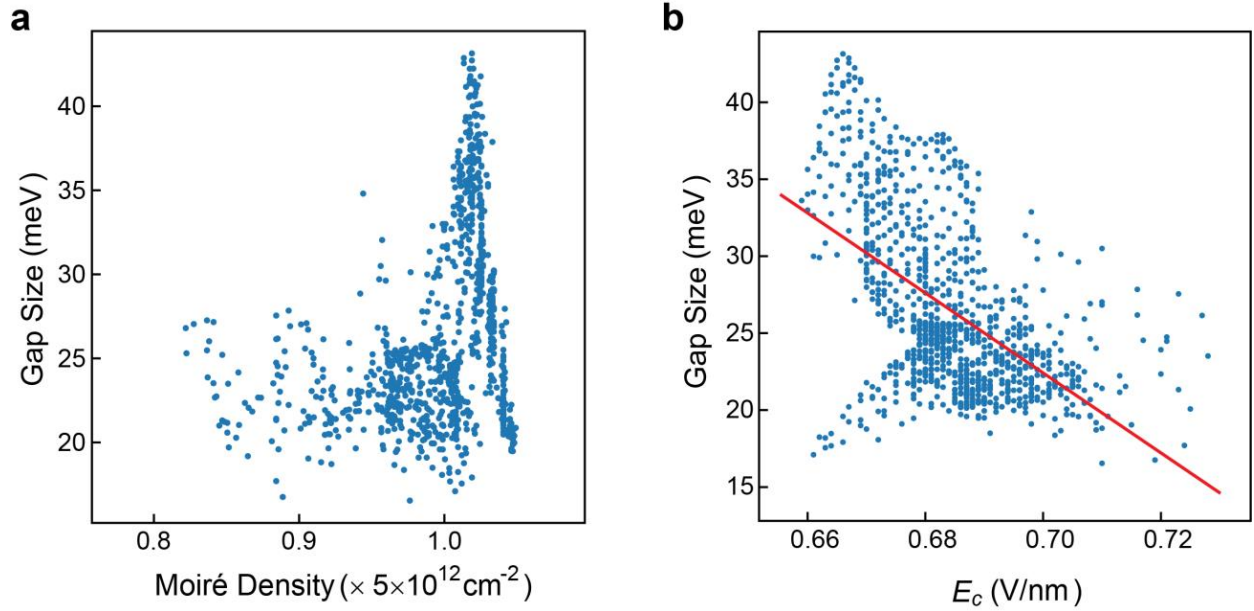

**Extended Data Figure 4 | Correlation between the charge gap size, the moiré density and the critical electric field. a,** Correlation between the Mott gap size (at $E = 0.56$ V/nm) and the moiré density as determined from the spatial maps in Fig. 4d,e in the main text. No obvious correlation is observed. **b,** Correlation between the Mott gap size and the critical electric field as determined from the spatial maps in Fig. 4e,f in the main text. A negative correlation is observed as shown by the linear fit.

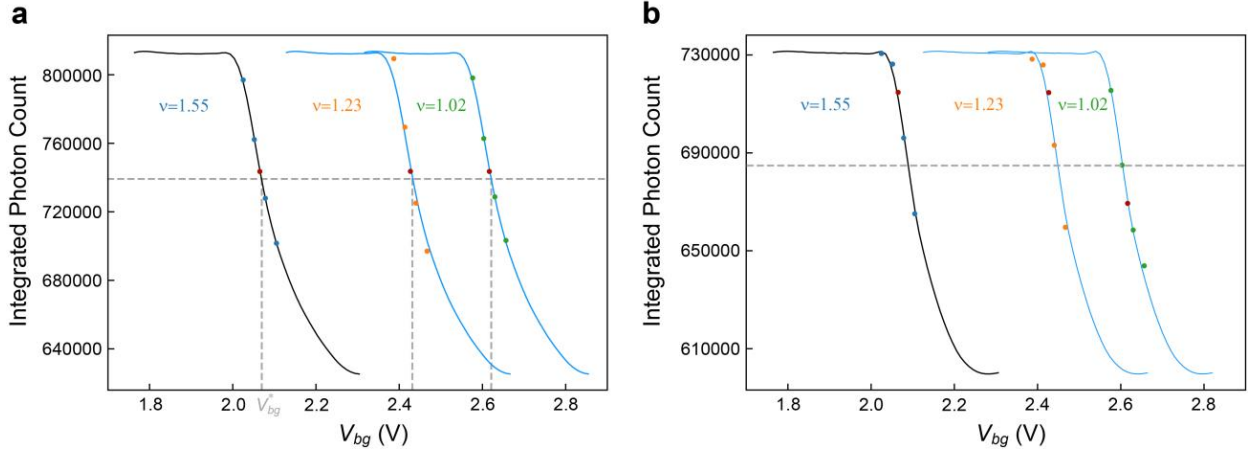

**Extended Data Figure 5 | Extraction of $\mu$ in imaging experiment. a,b,** Examples of extracting $\mu$ (at $E = 0.56$ V/nm) at the representative pixel 'R' (**a**) and at a different sample location (**b**). The black curves are the spatial dependent characteristic response curves. At each $\nu$ and $E$, we kept 'R' at the set point through a feedback loop; the reflection count at each pixel was recorded (red dots). We then recorded the reflection counts at four additional bottom gate voltages near the set point (blue, orange and green dots). The characteristic response curves are then shifted horizontally to match to the five data points for each $\nu$. The dashed lines correspond to 60 % reflection contrast of the response curves. At each $\nu$, we use the bottom gate voltage at the intersection of the dashed line and the response curve to obtain $\mu$.

19